\documentclass[epj]{svjour}
\usepackage{graphics, amssymb, amsfonts, amsmath}
\begin{document}
\title{Preparation of entanglement between atoms in spatially separated cavities via fiber loss}
\author{Shi-Lei Su\inst{1} \and Xiao-Qiang Shao\inst{2} \and Qi Guo\inst{1} \and Liu-Yong Cheng\inst{1} \and Hong-Fu Wang\inst{3} \and Shou Zhang\inst{1,3} 
\thanks{\emph{Corresponding author: szhang@ybu.edu.cn}}%
}                     
\institute{Department of Physics, Harbin Institute of Technology,
Harbin 150001, China \and Center for Quantum Sciences and School of Physics,
Northeast Normal University, Changchun 130024, China \and Department of Physics, College
of Science, Yanbian University, Yanji, Jilin 133002, China}
\date{Received: date / Revised version: date}
%
\abstract{
We propose a scheme to prepare a maximally entangled
state for two $\Lambda$-type atoms trapped in separate optical cavities coupled through an optical fiber
based on the combined effect of the unitary dynamics and the dissipative process.
Our work shows that the fiber loss, as well as the atomic spontaneous emission and the cavity
decay, is no longer undesirable, but requisite to prepare the distributed entanglement, which is meaningful for
the long distance quantum information processing tasks.
Originating from an arbitrary state, the desired state could be prepared without precise time control.
The robustness of the scheme is numerically demonstrated by considering various parameters.
\PACS{
      {03.67.Bg}{Entanglement production and manipulation}   \and
      {42.50.Pq}{Cavity quantum electrodynamics; micromasers }\and
      {03.65.Yz}{ Decoherence; open systems; quantum statistical methods}
     } 
} 
\maketitle
\section{Introduction}
Quantum entanglement~\cite{w001,w002} is a fundamental and fascinating quantum effect and
has been a key resource for quantum information science~(QIS)~\cite{w003,w004,w005}.
How to effectively prepare entanglement influences the development of QIS.
One traditional obstacle for preparing entanglement is the dissipation that induced by the inevitable
coupling between the quantum system and the environment, which would degrade the quantum coherence
of the system. The usual methods to deal with the dissipation can be divided into the following categories:
quantum error correction~\cite{w006,w007,w008}, decoherence-free subspace~\cite{w009,w010,w011},
geometric phase~\cite{w012,w013,w014,w015}, quantum-control technique~\cite{w016,w017,w018} and dissipative dynamics~
\cite{w019,w020,w021,w022,w023,w024,w025,w026,w027,w028,w029,w030,w031,w032,w033,w034,w035,w036,w037,w038,w039,w040,w043,w044,w045,w048,b048,w041,w042,w046,w047,b049,b050,nb051}. Compared with the other methods, dissipative dynamics has
unique characters since the dissipation is used to prepare entanglement so that the preparation process is
robust against decoherence. In particular, Kastoryano \emph{et al}. considered a dissipative
scheme for preparing a maximally entangled state of two $\Lambda$-type atoms trapped in one optical cavity~\cite{w030},
whose performance is better than that based on the unitary dynamics. Subsequently, this scheme was generalized to
coupled cavity system~\cite{w035}, atom-cavity-fiber system~\cite{w037} and the circuit quantum
electrodynamics system~\cite{w044}. Besides, Zheng \emph{et al}. proposed two schemes to
prepare the maximal entanglement between two atoms coupled to a decaying resonator~\cite{w048,b048}.
The common place of these schemes is that the cavity decay is used as resource
for state preparation. On the other hand, Rao and M{\o}lmer, Carr and Saffman considered two schemes~\cite{w041,w042}
to prepare the entanglement of Rydberg atoms via dissipation, respectively. And Shao \emph{et al}. gave two
dissipative schemes for three-dimensional entanglement
preparation~\cite{w046,w047}. These schemes demonstrated that the atomic spontaneous emission is quite capable of being the powerful
resource for entanglement preparation. Moreover, the schemes proposed in Refs.~\cite{w031,b049} indicate that the atomic spontaneous emission and
cavity decay could be used simultaneously to prepare the desired entanglement for two atoms in one cavity.

To realize the long distance and large-scale quantum information processing,
atom-cavity-fiber system is proposed~\cite{w049,w050} and has been an excellent platform
for distributed quantum computation~\cite{w051,w052}, quantum entanglement preparation
~\cite{w053,w054,w055,w056,w057,w058}, and quantum communication~\cite{w059}.
For these unitary-dynamics-based schemes, atomic spontaneous emission, cavity decay and the fiber loss are three main and undesirable
dissipative factors, which would affect the practical applications of the scheme in the QIS.

The previous dissipative schemes showed that the atomic spontaneous emission or cavity decay is no longer undesirable, but necessary for
maximal entanglement preparation for two atoms trapped in one cavity.
In this paper, we generalize the idea to the case when two atoms trapped into two distributed cavities that connected through an optical fiber.
Interestingly, we find that the fiber loss, like the other two dissipative factors, could also be used
to prepare the distributed entanglement, which is meaningful for the long distance and large-scale quantum information processing tasks.
The present work has the following features: (i) It has no specific requirement of the
initial state. (ii) It does not need to control
evolution time accurately. And, it is robustness on parameter
fluctuations. (iii) Different with the unitary-dynamics-based schemes in the atom-cavity-fiber system,
the present one shows that, the fiber loss, as well as the atomic spontaneous emission and the cavity decay,
is no longer undesirable, but requisite for the preparation of the distributed entanglement.
(iv) In contrast with the schemes based on the similar system~\cite{w037,w038}, the current one is special in the sense that
the decay of cavity and fiber modes are considered uncorrelated since we do not define the normal bosonic mode $\hat{c}$ to rewrite the Hamiltonian.

\section{Fundamental model}

\begin{figure}
\resizebox{0.5\textwidth}{!}{%
  \includegraphics{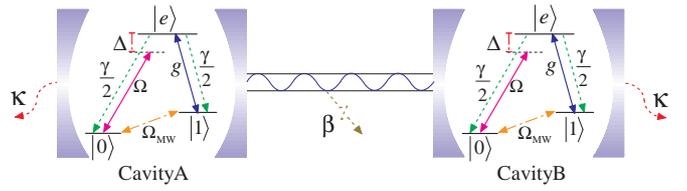}
}
\caption{Experimental setup and level diagram of $\Lambda$-type atoms.
  $\gamma$, $\kappa$ and $\beta$ denote the atomic spontaneous emission rate,
  cavity decay rate and fiber loss rate, respectively. Here, we assume the cavity decay rate
  of the two cavities are the same, and the excited state $|e\rangle$ spontaneously decays into
  two ground states with branching rate $\gamma/2$.}
\label{f001}       
\end{figure}

We consider the setup described in Fig.~\ref{f001}, where two
identical $\Lambda$-type atoms are individually trapped into two single mode
cavities coupled through an optical fiber with length \emph{l}. Each atom has two ground states $|0\rangle$ and
$|1\rangle$ and one excited state $|e\rangle$ with the
corresponding energies $\omega_{0}$, $\omega_{1}$ and
$\omega_{e}$, respectively. The atomic transition $|1\rangle\leftrightarrow|e\rangle$ is coupled resonantly to the
corresponding cavity mode with the coupling
constant $g$. Besides, the transition $|0\rangle\leftrightarrow|e\rangle$ is driven by the corresponding off-resonance optical laser with
detuning $\Delta$, and Rabi frequency $\Omega$. The transition between two ground states
$|0\rangle$ and $|1\rangle$ is coupled resonantly by means of a
microwave field with Rabi frequency $\Omega_{\rm MW}$.
In the short fiber limit $(l\bar{v})/(2\pi c)\leqslant1$, where $\bar{v}$ is the decay rate of the cavity fields into
a continuum of fiber modes, only one fiber mode interacts with the cavity mode~\cite{w050,w051,w052}. For simplicity, we assume the
interaction between cavity mode and fiber mode is resonant.
Under the rotating wave approximation, the Hamiltonian of the whole system could be written as (setting $\hbar=1$)
$H = H_{a,c,f} + H_{cl} + H_{mw}$,
\begin{eqnarray}\label{e001}
    H_{a,c,f} &=& \sum_{i={\rm A,B}}\Big[\sum_{j=0,1,e}\omega_{j}|j\rangle_{ii}\langle j|
     + \omega_{a}a^{\dag}_{i}a_{i}
     \cr\cr&& +~~ (g|e\rangle_{ii}\langle 1|a_{i} + {\rm H.c.})\Big]
     \cr\cr&& +~~ \omega_{b}b^{\dag}b
     + \big[\nu b(a^{\dag}_{A} + a^{\dag}_{B}) + {\rm H.c.}\big],
\end{eqnarray}
\begin{equation}\label{e002}
    H_{cl} = \Omega\sum_{i={\rm A,B}} e^{i\omega t}|0\rangle_{ii}\langle e| + {\rm H.c.},
\end{equation}
\begin{equation}\label{e003}
    H_{mw} = \Omega_{\rm MW}e^{i\omega_{\rm MW}t}(|0\rangle_{\rm AA}\langle 1| - |0\rangle_{\rm BB}\langle 1|) + {\rm H.c.},
\end{equation}
in which $a_{i}$ and $a^{\dag}_{i}$ denote the annihilation and creation
operators for the optical mode of cavity \emph{i}, respectively. $b$ and $b^{\dag}$ denote the annihilation
and creation operator for the fiber mode, respectively. $\omega_{a}$ and $\omega_{b}$ denote the frequencies of
cavity mode and the fiber mode, respectively. $\omega$ and $\omega_{\rm MW}$ stand for the frequencies of the classical laser
field and the microwave field, respectively. $\nu$ is the coupling strength between the cavity mode and the fiber mode.
Then, we define the excitation number operator of the total system
$N_{e}=\sum_{i=\rm A,B}(|e\rangle_{ii}\langle e| + a^{\dag}_{i}a_{i}) + b^{\dag}b$.
Under the weak excitation condition and if the
initial state is in the zero excitation subspace, we can safely discard the
subspace with excitation number greater than or equals two.

With the dissipation being included, the dynamics of the system in Lindblad form could be
described by the master equation
\begin{equation}\label{e009}
\dot{\rho} = i[\rho, H] + \sum_{j}\Big[L_{j}\rho\L_{j}^{\dag} - \frac{1}{2}(\L_{j}^{\dag}L_{j}\rho +\rho\L_{j}^{\dag}L_{j}) \Big],
\end{equation}
where $L_{j}$ is the so-called Lindblad operators governing dissipation. Specifically, in the current scheme the Lindblad operators can be expressed as $L_{\beta}=\sqrt{\beta}b$, $L_{\gamma_{\rm 1}}=\sqrt{\gamma/2}|0\rangle_{\rm AA}\langle e|$,
$L_{\gamma_{\rm 2}}=\sqrt{\gamma/2}|1\rangle_{\rm AA}\langle e|$, $L_{\gamma_{3}}=\sqrt{\gamma/2}|0\rangle_{\rm BB}\langle e|$,
$L_{\gamma_{4}}=\sqrt{\gamma/2}|1\rangle_{\rm BB}\langle e|$, $L_{\kappa_{1}}=\sqrt{\kappa}a_{\rm A}$ and $L_{\kappa_{2}}=\sqrt{\kappa}a_{\rm B}$.
$L_{\beta}$ describes the dissipation induced by the fiber loss. $L_{\gamma_{\rm 1}}$ and $L_{\gamma_{\rm 2}}$ describe the dissipation induced
by the spontaneous emission of atom in cavity A. $L_{\gamma_{\rm 3}}$ and $L_{\gamma_{\rm 4}}$ describe the dissipation induced
by the spontaneous emission of atom in cavity B. $L_{\kappa_{1}}=\sqrt{\kappa}a_{\rm A}$ and $L_{\kappa_{2}}=\sqrt{\kappa}a_{\rm B}$ describe
the dissipation induced by the leakage of cavities A and B, respectively. Since cavity A and cavity B are distant from each other, the dissipation
processes are spatially separated.

\section{Preparation of the distributed entanglement}      

\subsection{Dressed states}
To see clearly the roles of the classical laser field, microwave field and dissipative factors, we identify the eigenstates
of the Hamiltonian $H_{a,c,f}$ in zero and one excitation subspace and use them as dressed states.
\begin{table}
\caption{The eigenstates and the corresponding eigenenergies of Hamiltonian $H_{a,c,f}$ in zero excitation subspaces.
Here, $|\alpha\beta\rangle|\gamma\delta\epsilon\rangle$ represents
that atom in cavity A~(B) is in the state $|\alpha\rangle$~($|\beta\rangle$),
optical mode of the cavity A~(B) is in the state $|\gamma\rangle~(|\epsilon\rangle)$
and of the fiber is in state $|\delta\rangle$.}
\label{t001}       
\begin{tabular}{ll}
\hline\hline\noalign{\smallskip}
Eigenstate & Eigenenergy\\
  \hline
$|00\rangle|000\rangle$ & ~0\\
$|S\rangle|000\rangle$ & ~$\omega_{1}$ \\
$|T\rangle|000\rangle$ & ~$\omega_{1}$ \\
$|11\rangle|000\rangle$ & 2$\omega_{1}$\\
\noalign{\smallskip}\hline\hline
\end{tabular}
\end{table}

\begin{table}
\caption{The eigenstates and the corresponding eigenenergies of Hamiltonian $H_{a,c,f}$ in one excitation subspace.}
\label{t002}       
\begin{tabular}{ll}
\hline\hline\noalign{\smallskip}
Eigenstate & Eigenenergy\\
  \hline
$|\varphi\rangle_{1}$ & $\omega_{e}-\omega_{1}$\\
$|\varphi\rangle_{2}$ & $\omega_{e}-\omega_{1}-\sqrt{2}\nu$\\
$|\varphi\rangle_{3}$ & $\omega_{e}-\omega_{1}+\sqrt{2}\nu$\\
$|\varphi\rangle_{4}$ & $\omega_{e}+\omega_{1}$\\
$|\varphi\rangle_{5}$ & $\omega_{e}+\omega_{1}-g$\\
$|\varphi\rangle_{6}$ & $\omega_{e}+\omega_{1}+g$\\
$|\varphi\rangle_{7}$ & $\omega_{e}+\omega_{1}-g_{1}$\\
$|\varphi\rangle_{8}$ & $\omega_{e}+\omega_{1}+g_{1}$\\
$|T_{1}\rangle$, $|S_{1}\rangle$ & $\omega_{e}-\sqrt{g_{1}^{2}-g_{3}^{2}}/\sqrt{2}$ \\
$|T_{2}\rangle$, $|S_{2}\rangle$& $\omega_{e}+\sqrt{g_{1}^{2}-g_{3}^{2}}/\sqrt{2}$ \\
$|T_{3}\rangle$, $|S_{3}\rangle$ & $\omega_{e}-\sqrt{g_{1}^{2}+g_{3}^{2}}/\sqrt{2}$ \\
$|T_{4}\rangle$, $|S_{4}\rangle$ & $\omega_{e}+\sqrt{g_{1}^{2}+g_{3}^{2}}/\sqrt{2}$ \\
\noalign{\smallskip}\hline\hline
\end{tabular}
\end{table}
In Tables~\ref{t001} and \ref{t002}, we show the eigenstates and the
corresponding eigenvalues (setting $\omega_{0}$=0) with the notations shown in \textbf{Appendix},
and $|T\rangle$ is the maximal entanglement we want to prepare for the two distributed atoms.

\subsection{Roles of the microwave field and the classical laser field}\label{s302}

\begin{figure}
  \resizebox{0.5\textwidth}{!}{%
  \includegraphics{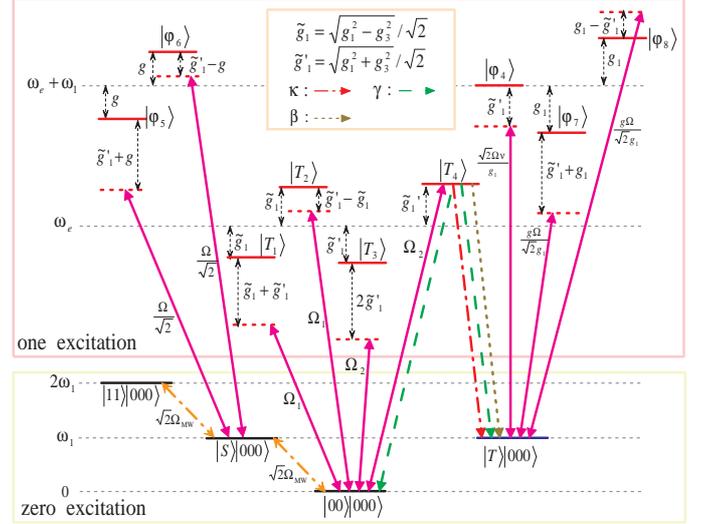}}
  \caption{Level configuration and the transitions of the states in the dressed state picture. Microwave field causes resonant transitions among states
  $|11\rangle|000\rangle$, $|S\rangle|000\rangle$ and $|00\rangle|000\rangle$. Classical laser field causes resonant transitions between
  states $|00\rangle|000\rangle$ and $|T_{4}\rangle$. The other transitions between state in zero excitation subspace and that in one excitation caused
  by classical laser field are non-resonant with different detunings. Dissipative factors, i.e., atomic spontaneous emission, cavity decay and fiber loss, are critical
  to translate the state $|T_{4}\rangle$ into the desired state. We use the notations $\Omega_{1} = \sqrt{2}g\nu\Omega/\sqrt{g_{3}^{4}+g^{2}g_{3}^{2}-2\nu^{2}g_{3}^{2}}$ and
 $\Omega_{2}=\sqrt{2}g\nu\Omega/\sqrt{g_{3}^{4} - g^{2}g_{3}^{2}+2\nu^{2}g_{3}^{2}}$ for simplicity. }\label{f002}
\end{figure}
Under the dressed state picture, Hamiltonian $H_{mw}$ can be rewritten as
\begin{eqnarray}\label{e005}
   H_{mw} &=& \sqrt{2}~\Omega_{\rm MW}~e^{i\omega_{\rm MW}t}(|S\rangle|000\rangle\langle000|\langle11|
   \cr\cr&&-|00\rangle|000\rangle\langle000|\langle S|)
  {\rm + ~H.c.}.
\end{eqnarray}
Similarly, $H_{cl}$ can be rewritten as
\begin{eqnarray}\label{e006}
   H_{cl} &=& \frac{\sqrt{2}g\nu\Omega}{\sqrt{g_{3}^{4}+g^{2}g_{3}^{2}-2\nu^{2}g_{3}^{2}}}e^{i\omega t}|00\rangle|000\rangle(\langle T_{1}| + \langle T_{2}|)
   \cr\cr&& + \frac{\sqrt{2}g\nu\Omega}{\sqrt{g_{3}^{4} - g^{2}g_{3}^{2}+2\nu^{2}g_{3}^{2}}}e^{i\omega t}|00\rangle|000\rangle(\langle T_{3}| + \langle T_{4}|)
   \cr\cr&&+\frac{\sqrt{2}\Omega\nu}{g_{1}}e^{i\omega t}|T\rangle|000\rangle\langle\varphi_{4}|
   \cr\cr&&+\frac{g\Omega}{\sqrt{2}g_{1}}e^{i\omega t}|T\rangle|000\rangle(\langle\varphi_{7}| + \langle\varphi_{8}|)
   \cr\cr&&+\frac{\Omega}{\sqrt{2}}e^{i\omega t}|S\rangle|000\rangle(\langle\varphi_{5}| + \langle\varphi_{6}|) + {\rm H.c.}.
\end{eqnarray}
In the interaction picture with respect to the Hamiltonian $H_{a,c,f}$ that
expressed by the eigenvectors and eigenvalues in zero and one excitation subspace,
Eqs.~(\ref{e005}) and (\ref{e006}) can be transformed to
\begin{eqnarray}\label{e007}
   H_{mw} &=& \sqrt{2}~\Omega_{\rm MW}(|S\rangle|000\rangle\langle000|\langle11|-|00\rangle|000\rangle\langle000|\langle S|)
  \cr\cr&&{\rm + ~H.c.}
\end{eqnarray}
and
\begin{eqnarray}\label{e008}
   H_{cl} &=& \frac{\sqrt{2}g\nu\Omega}{\sqrt{g_{3}^{4}+g^{2}g_{3}^{2}-2\nu^{2}g_{3}^{2}}}\Big[e^{i(\omega-\omega_{e}+\sqrt{g_{1}^{2}-g_{3}^{2}}/\sqrt{2}) t}\cr\cr&&\times|00\rangle|000\rangle\langle T_{1}| + e^{i(\omega-\omega_{e}-\sqrt{g_{1}^{2}-g_{3}^{2}}/\sqrt{2}) t}|00\rangle|000\rangle\langle T_{2}|\Big]
   \cr\cr&& +\frac{\sqrt{2}g\nu\Omega}{\sqrt{g_{3}^{4} - g^{2}g_{3}^{2}+2\nu^{2}g_{3}^{2}}}\Big[e^{i(\omega-\omega_{e}+\sqrt{g_{1}^{2}+g_{3}^{2}}/\sqrt{2}) t}\cr\cr&&\times|00\rangle|000\rangle\langle T_{3}| + e^{i(\omega-\omega_{e}-\sqrt{g_{1}^{2}+g_{3}^{2}}/\sqrt{2}) t}|00\rangle|000\rangle\langle T_{4}|\Big]
   \cr\cr&& +\frac{\sqrt{2}\Omega\nu}{g_{1}}e^{i(\omega-\omega_{e}) t}|T\rangle|000\rangle\langle\varphi_{4}|
    \cr\cr&&+\frac{g\Omega}{\sqrt{2}g_{1}}\Big[e^{i(\omega-\omega_{e}+g_{1}) t}|T\rangle|000\rangle\langle\varphi_{7}|\cr\cr&& + e^{i(\omega-\omega_{e}-g_{1}) t}|T\rangle|000\rangle\langle\varphi_{8}|\Big]
   \cr\cr&& +\frac{\Omega}{\sqrt{2}}\Big[e^{i(\omega-\omega_{e}+g) t}|S\rangle|000\rangle\langle\varphi_{5}| \cr\cr&&+ e^{i(\omega-\omega_{e}-g) t}|S\rangle|000\rangle\langle\varphi_{6}|\Big] + {\rm H.c.}.
\end{eqnarray}
Equation.~(\ref{e007}) shows that Hamiltonian $H_{mw}$ causes resonant transitions among states $|11\rangle|000\rangle$, $|S\rangle|000\rangle$ and $|00\rangle|000\rangle$, as we shall see later, which is exactly the reason why the scheme is independent of the initial states.
From Eq.~(\ref{e008}), one can see that classical laser field causes interactions between the states
in zero excitation subspace and that in one excitation subspace, and the detuning of the interaction
could be adjusted through choosing the values of $\omega$, $\omega_{e}$, $g$ and $\nu$ according to the requirement of the scheme.
For example, if one choose $\omega_{e}-\omega=-\sqrt{g_{1}^{2}+g_{3}^{2}}/\sqrt{2}$, $|00\rangle|000\rangle$ would couple resonantly to
$|T_{4}\rangle$, while other terms in Eq.~(\ref{e008}) would
undergo non-resonant interactions with different detunings.

\subsection{Roles of the dissipative factors}

Dissipation, which can occur via the fiber loss, atomic spontaneous emission, and cavity decay,
is an integrant component of the current state preparation scheme. The states in one excitation
subspace would be transformed to the corresponding states in zero excitation subspace via dissipation.
Interestingly, it is easy to find that the second term of $|T_{4}\rangle$ would be transformed to the state
$|T\rangle|000\rangle$ (have not considered the normalization factor) when $L_{\beta}$ works.
The first and the third terms of $|T_{4}\rangle$ would be transformed to the state
$|T\rangle|000\rangle$ when $L_{\kappa_{1}}$ and $L_{\kappa_{2}}$ work. Besides, the fourth term of $|T_{4}\rangle$
would also be converted to $|T\rangle|000\rangle$ when $L_{\gamma_{2}}$ and $L_{\gamma_{4}}$ work.
Since $|T\rangle|000\rangle$ is the product
state of the atomic maximally entangled state $|T\rangle$, the cavity mode vacuum state, and the fiber mode vacuum state,
the scheme would be considered successful if $|T\rangle|000\rangle$ is prepared.
Therefore, in order to prepare the desired state, it is better if other undesired states in zero excitation subspace could be coupled resonantly to
$|T_{4}\rangle$ in one excitation subspace directly or indirectly.

\subsection{Preparation process}

\begin{figure}
\begin{center}
  \resizebox{0.5\textwidth}{!}{%
  \includegraphics{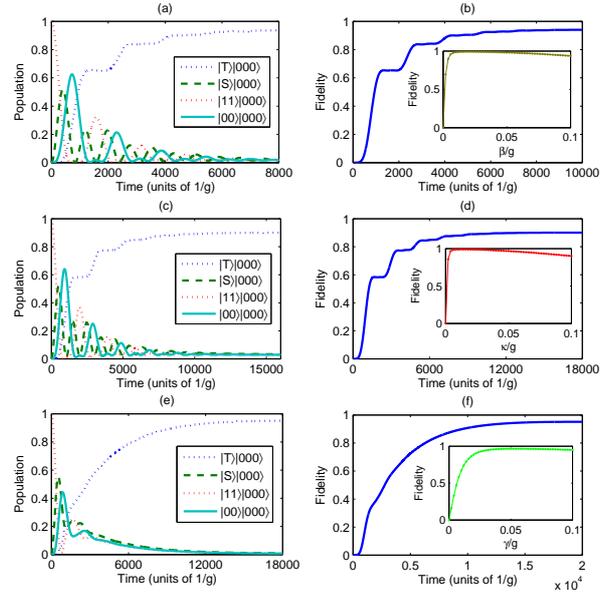}}
  \caption{(a)(c)(e) Populations of the states in zero subspace with the initial state $|11\rangle|000\rangle$. (b)(d)(f) Fidelities of the desired state with the initial state $|11\rangle|000\rangle$.
  In subfigures~(a) and (b), the parameters are chosen as $\Omega = 0.008g, \Omega_{\rm MW} = 0.25\Omega, \nu = g$. And the dissipative factors are chosen as $\beta=0.1g$, $\kappa = 0$, and $\gamma=0$.
  The inset of subfigure~(b) is plotted with the dissipative factors $\kappa = 0$, $\gamma=0$, and $\beta$ varies from 0 to $0.1g$ at the time $8\times10^3/g$.
  In subfigures~(c) and (d), the parameters are chosen as $\Omega = 0.008g, \Omega_{\rm MW} = 0.2\Omega, \nu = g$. And the dissipative factors are chosen as $\beta=0$, $\kappa = 0.1g$, and $\gamma=0$.
  The inset of subfigure~(d) is plotted with the dissipative factors $\beta = 0$, $\gamma=0$, and $\kappa$ varies from 0 to $0.1g$ at the time $1.6\times10^4/g$.
  In subfigures~(e) and (f), the parameters are chosen as $\Omega = 0.008g, \Omega_{\rm MW} = 0.2\Omega, \nu = g$. And the dissipative factors are chosen as $\beta=0$, $\kappa = 0$, and $\gamma=0.1g$.
  The inset of subfigure~(f) is plotted with the dissipative factors $\beta = 0$, $\kappa=0$, and $\gamma$ varies from 0 to $0.1g$ at the time $1.6\times10^4/g$.}\label{f003}
\end{center}
\end{figure}

If the parameters satisfy $\omega_{e}-\omega=-\sqrt{g_{1}^{2}+g_{3}^{2}}/\sqrt{2}$, $|00\rangle|000\rangle$ would couple resonantly to
$|T_{4}\rangle$ while the other terms in Eq.~(\ref{e008}) underdo non-resonant interactions with different detunings.
First, we take into account the case that the initial state is $|11\rangle|000\rangle$. As shown in Sec.~\ref{s302}, $|11\rangle|000\rangle$
would be transformed to $|00\rangle|000\rangle$ through the intermediate state $|S\rangle|000\rangle$.
Since $|00\rangle|000\rangle$ couples resonantly to $|T_{4}\rangle$,
the initial state $|11\rangle|000\rangle$ would be transformed to
$|T\rangle|000\rangle$ via dissipation finally. The process could be expressed as
$~~~~|11\rangle|000\rangle\xrightarrow[]{H_{mw}}|S\rangle|000\rangle\xrightarrow[]{H_{mw}}
|00\rangle|000\rangle\xrightarrow[]{H_{cl}}|T_{4}\rangle\xrightarrow[]{\gamma, \kappa, \beta}|T\rangle|000\rangle$.
Similarly, if the initial state is $|S\rangle|000\rangle$ or $|00\rangle|000\rangle$, it would also be converted to the desired state finally.
Besides, since $|01\rangle|000\rangle$ or $|10\rangle|000\rangle$ could be regarded as a superposition of the states
$|S\rangle|000\rangle$ and $|T\rangle|000\rangle$, and $|S\rangle|000\rangle$ would be transformed to $|T\rangle|000\rangle$ finally, the
desired state could also be achieved if the initial state is $|01\rangle|000\rangle$ or $|10\rangle|000\rangle$.
In brief, if the initial state is $|T\rangle|000\rangle$, it is stable and keeps invariant. Otherwise, it would be transformed to $|T\rangle|000\rangle$ via
unitary dynamics and dissipative process. As a result, the population of $|T\rangle|000\rangle$ accumulates as time grows.

In the above analysis, for simplicity we only consider the dissipative channel $|e\rangle\rightarrow|1\rangle$ when we study the roles of the atomic spontaneous emission.
Nevertheless, the atomic spontaneous emission has another possibilities, $|e\rangle\rightarrow|0\rangle$. By the time that happens, the fourth term of the state $|T_{4}\rangle$ would be converted to
$|00\rangle|000\rangle$, which is coupled resonantly to the state $|T_{4}\rangle$ through the classical laser field and thus has little or no effect on
the state preparation.
The detailed level configuration and the transformed process of the whole scheme are shown in Fig.~\ref{f002}.

We also noticed that $|T_{1}\rangle$, $|T_{2}\rangle$, and $|T_{3}\rangle$ could also be transformed to the desired state through the dissipative process.
Thus, if the selection of $\omega_{e}-\omega$ could realize the resonant interactions between the state $|00\rangle|000\rangle$ and any one of the three states $|T_{1}\rangle$, $|T_{2}\rangle$ and $|T_{3}\rangle$, the desired state could also be prepared via the dissipation.

\section{Discussion}

\begin{figure}
\resizebox{0.5\textwidth}{!}{%
  \includegraphics{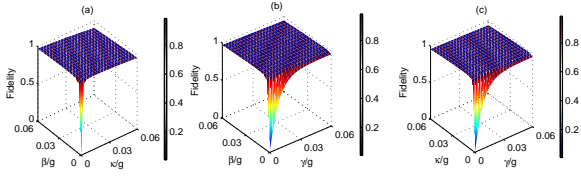}}
  \caption{Fidelity of the desired state with the initial state $|11\rangle|000\rangle$ at the time $1.5\times10^4/g$. The parameters are chosen as $\Omega = 0.008g, \Omega_{\rm MW} = 0.25\Omega$, and $\nu = g$. (a) The dissipative factors $\beta$ and $\kappa$ vary from 0 to $0.06g$, and $\gamma$ is set to zero. (b) The dissipative factors $\beta$ and $\gamma$ vary from 0 to $0.06g$, and $\kappa$ is set to zero. (c) The dissipative factors $\gamma$ and $\kappa$ vary from 0 to $0.06g$, and $\beta$ is set to zero.}\label{f004}
\end{figure}
\begin{figure}
\resizebox{0.55\textwidth}{!}{%
  \includegraphics{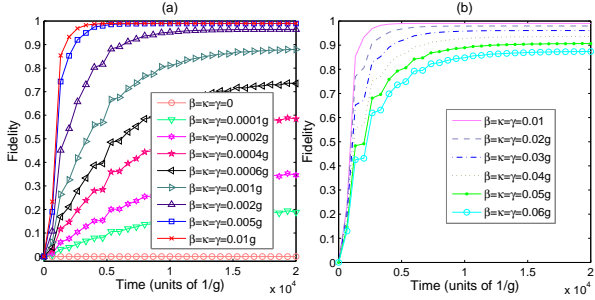}}
  \caption{Fidelity of the desired state with the initial state $|11\rangle|000\rangle$. The parameters are chosen as $\Omega = 0.008g, \Omega_{\rm MW} = 0.25\Omega$, and $\nu = g$. The dissipative factors are set as $\beta=\kappa=\gamma$.}\label{f005}
\end{figure}
\begin{figure}
\resizebox{0.5\textwidth}{!}{%
  \includegraphics{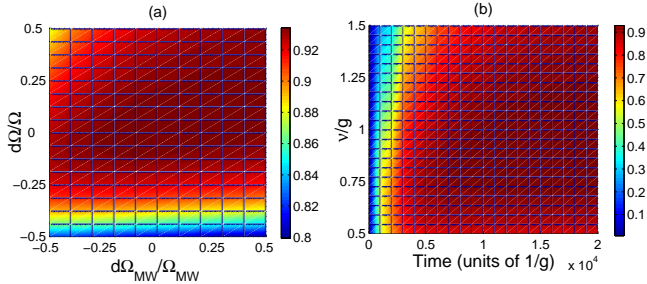}}
  \caption{(a) Fidelity of the desired state versus the relative fluctuations of $\Omega$ and $\Omega_{\rm MW}$ with the initial state $|11\rangle|000\rangle$
  at the time $2\times10^{4}/g$. The notations d$\Omega$ and d$\Omega_{\rm MW}$ in the axis labels denote the deviations of $\Omega$ and $\Omega_{\rm MW}$, respectively. The parameters are chosen as $\Omega = 0.008g, \Omega_{\rm MW} = 0.002g$, and $\nu = g$. (b) Fidelity of the desired state versus $\nu$ and evolution time with the initial state $|11\rangle|000\rangle$. The parameters are chosen as $\Omega = 0.008g$ and $\Omega_{\rm MW} = 0.002g$. Both figures are plotted with the dissipative factors $\beta=\kappa=\gamma=0.04g$.}\label{f006}
\end{figure}
To verify the feasibility of the scheme, we solve the master equation numerically in zero and one excitation subspace.
For the purpose of studying the effect of each of the dissipative factors, we would first consider one factor at a time. Then, the combined effect
of the dissipative factors would be considered. In Figs.~\ref{f003}(a) and~\ref{f003}(b), we plot the populations and the fidelity of the scheme when the dissipative factors satisfy $\beta=0.1g$, $\kappa = 0$, and $\gamma=0$. The results show
that the desired state could be prepared with the fidelity more than 94\% when the evolution time equals $10^{4}/g$. The inset in Fig.~\ref{f003}(b) shows that
when $\kappa=\gamma=0$, the fiber loss is critical for the state preparation since the fidelity would be zero if the fiber loss is not exist.
Similarly, numerical results in Figs.~\ref{f003}(c) and \ref{f003}(d), and ~\ref{f003}(e) and \ref{f003}(f) indicate that the cavity decay
and atomic spontaneous emission could also be utilized as resources
to prepare the maximal entanglement with high fidelity, respectively.
In Fig.~\ref{f004}, we plot the fidelity when two out of three dissipative factors are considered at the same time. The result gives a further verification that each of the dissipative factors could be utilized to prepare the distributed entanglement.

In Fig.~\ref{f005}, we plot the fidelity when the three dissipative factors are considered at the same time. One can see from Fig.~\ref{f005}(a) that the fidelity increases as the dissipative factors increases. However, as shown in Fig.~\ref{f005}(b), further increase of the dissipative factors would decrease the
performance of the scheme. In fact, like most of the dissipative schemes, the present one is based on the combined effect of the unitary dynamics and the dissipative process. When the values of the dissipative factors are set to zero, the scheme would not succeed. As the dissipative factors increases,
the role of the dissipative process becomes evident and the fidelity increases. Nevertheless, further increase of the dissipative factors
would lead to an adverse effect on the unitary dynamics and thus decrease the overall performance of the scheme.
Besides, in Fig.~\ref{f006}(a) we plot the fidelity versus the fluctuations of $\Omega$ and $\Omega_{\rm MW}$, which shows that the scheme is robust on the variation of $\Omega$ and $\Omega_{\rm MW}$ since the fidelity remains higher than 80\% even when the relative errors of these two factors reach 50\% simultaneously. And Fig.~\ref{f006}(b) indicates that the scheme is robust on the variation of the coupling strength between the cavity mode and the fiber mode.

If we use the experimental cavity-parameters $(g,\kappa,\gamma)\sim2\pi\times(34,4.1,3.6)$MHz~\cite{b060} and select $\Omega = 0.015g, \Omega_{\rm MW} = 0.36\Omega, \beta \simeq \kappa$, and $\nu = 0.9g$ for simulation, the fidelity is about 71.7\%. And this value may further decrease
 after considering various realistic experimental conditions. Nevertheless, as shown in Fig.~\ref{f005}(b), the performance can be greatly improved as the dissipative rate decreases within a certain range. Thus, a better cavity with high cooperativity should be satisfied for implementing the scheme
 experimentally. In addition, the short fiber limit $l\bar{\nu}/2\pi c<1$ should be fulfilled during the whole experimental process.
 On that premise, the scheme has certain requirements about the value of $\bar{\nu}$ for realizing long-distance quantum information processing.

In contrast with the unitary-dynamics-based schemes~\cite{w053,w054,w055,w056,w057,w058,w060}, the fiber loss takes on a new role since it constitutes a vital passage from the state in one excitation subspace to the desired state in the zero excitation subspace.
In comparison with the dissipative schemes~\cite{w037,w038} based on the similar system, the difference is that the current scheme has not introduced the normal bosonic modes. Therefore, the bosonic modes of the cavity and the fiber are considered to be completely uncorrelated in our scheme.

\section{Conclusion}

In conclusion, we have designed an alternative scheme to prepare the distributed entanglement in the atom-cavity-fiber system via the fiber loss as well as the atomic spontaneous emission and cavity decay. And the dissipation is an essential part of the scheme.
We hope that out work may be useful for the distributed quantum information processing tasks in the near future.
\newline\newline
\begin{flushleft}
   \textbf{Appendix}
\end{flushleft}

\begin{eqnarray}\label{e004}
    |S\rangle &\equiv& (|01\rangle-|10\rangle)/\sqrt{2},~~
    |T\rangle\equiv(|01\rangle+|10\rangle)/\sqrt{2},~~
    \cr\cr|\varphi\rangle_{1}&\equiv&(|00\rangle|100\rangle-|00\rangle|001\rangle)/\sqrt{2},
    \cr\cr|\varphi\rangle_{2}&\equiv&(|00\rangle|100\rangle+|00\rangle|001\rangle-\sqrt{2}|00\rangle|010\rangle)/2,~~
    \cr\cr|\varphi\rangle_{3}&\equiv&(|00\rangle|100\rangle+|00\rangle|001\rangle+\sqrt{2}|00\rangle|010\rangle)/2,
    \cr\cr|\varphi\rangle_{4}&\equiv&(|e1\rangle|000\rangle+|1e\rangle|000\rangle-\frac{g}{\nu}|11\rangle|010\rangle)/\sqrt{2+(\frac{g}{\nu})^{2}},
    \cr\cr|\varphi\rangle_{5}&\equiv&(|11\rangle|001\rangle+|e1\rangle|000\rangle-|1e\rangle|000\rangle-|11\rangle|100\rangle)/2,
    \cr\cr|\varphi\rangle_{6}&\equiv&(|11\rangle|100\rangle+|e1\rangle|000\rangle-|1e\rangle|000\rangle-|11\rangle|001\rangle)/2,
    \cr\cr|\varphi\rangle_{7}&\equiv&(|e1\rangle|000\rangle+|1e\rangle|000\rangle+\frac{2\nu}{g}|11\rangle|010\rangle
    -\frac{g_{1}}{g}|11\rangle|001\rangle\cr\cr&&-\frac{g_{1}}{g}|11\rangle|100\rangle)/\sqrt{2(\frac{g_{1}}{g})^{2}+(\frac{2\nu}{g})^{2}+2},
    \cr\cr|\varphi\rangle_{8}&\equiv&(|e1\rangle|000\rangle+|1e\rangle|000\rangle+\frac{2\nu}{g}|11\rangle|010\rangle
    +\frac{g_{1}}{g}|11\rangle|001\rangle\cr\cr&&+\frac{g_{1}}{g}|11\rangle|100\rangle)/\sqrt{2(\frac{g_{1}}{g})^{2}+(\frac{2\nu}{g})^{2}+2},
    \cr\cr|T_{1}\rangle&\equiv&\Big[\frac{\sqrt{g^{2}_{1}-g^{2}_{3}}~(g^{2}+g^{2}_{3})}{2\sqrt{2}g\nu^{2}}
    (|01\rangle|100\rangle+|10\rangle|001\rangle)\cr\cr&&-\frac{g^{2}_{2}+g^{2}_{3}}{2g\nu}(|01\rangle|010\rangle + |10\rangle|010\rangle)
    \cr\cr&&-\frac{\sqrt{g^{2}_{1}-g_{3}^{2}}}{\sqrt{2}g}(|01\rangle|001\rangle+ |10\rangle|100\rangle)\cr\cr&&+(|0e\rangle|000\rangle + |e0\rangle|000\rangle)\Big]/\frac{\sqrt{2(g_{3}^{4}+g^{2}g_{3}^{2}-2g_{3}^{2}\nu^{2})}}{g\nu},
    \cr\cr|S_{1}\rangle&\equiv&\Big[\frac{\sqrt{g^{2}_{1}-g^{2}_{3}}~(g^{2}+g^{2}_{3})}{2\sqrt{2}g\nu^{2}}
    (|01\rangle|100\rangle-|10\rangle|001\rangle)\cr\cr&&-\frac{g^{2}_{2}+g^{2}_{3}}{2g\nu}(|01\rangle|010\rangle - |10\rangle|010\rangle)
    \cr\cr&&-\frac{\sqrt{g^{2}_{1}-g_{3}^{2}}}{\sqrt{2}g}(|01\rangle|001\rangle - |10\rangle|100\rangle)\cr\cr&&+(|0e\rangle|000\rangle - |e0\rangle|000\rangle)\Big]/\frac{\sqrt{2(g_{3}^{4}+g^{2}g_{3}^{2}-2g_{3}^{2}\nu^{2})}}{g\nu},
    \cr\cr|T_{2}\rangle&\equiv&\Big[-\frac{\sqrt{g^{2}_{1}-g^{2}_{3}}~(g^{2}+g^{2}_{3})}{2\sqrt{2}g\nu^{2}}
    (|01\rangle|100\rangle+|10\rangle|001\rangle)\cr\cr&&-\frac{g^{2}_{2}+g^{2}_{3}}{2g\nu}(|01\rangle|010\rangle + |10\rangle|010\rangle)
    \cr\cr&&+\frac{\sqrt{g^{2}_{1}-g_{3}^{2}}}{\sqrt{2}g}(|01\rangle|001\rangle + |10\rangle|100\rangle)\cr\cr&&+(|0e\rangle|000\rangle + |e0\rangle|000\rangle)\Big]/\frac{\sqrt{2(g_{3}^{4}+g^{2}g_{3}^{2}-2g_{3}^{2}\nu^{2})}}{g\nu},
    \cr\cr|S_{2}\rangle&\equiv&\Big[-\frac{\sqrt{g^{2}_{1}-g^{2}_{3}}~(g^{2}+g^{2}_{3})}{2\sqrt{2}g\nu^{2}}
    (|01\rangle|100\rangle-|10\rangle|001\rangle)\cr\cr&&-\frac{g^{2}_{2}+g^{2}_{3}}{2g\nu}(|01\rangle|010\rangle - |10\rangle|010\rangle)
    \cr\cr&&+\frac{\sqrt{g^{2}_{1}-g_{3}^{2}}}{\sqrt{2}g}(|01\rangle|001\rangle - |10\rangle|100\rangle)\cr\cr&&+(|0e\rangle|000\rangle - |e0\rangle|000\rangle)\Big]/\frac{\sqrt{2(g_{3}^{4}+g^{2}g_{3}^{2}-2g_{3}^{2}\nu^{2})}}{g\nu},
    \cr\cr|T_{3}\rangle&\equiv&\Big[\frac{\sqrt{g^{2}_{1}+g^{2}_{3}}~(g^{2}-g^{2}_{3})}{2\sqrt{2}g\nu^{2}}
    (|01\rangle|100\rangle+|10\rangle|001\rangle)\cr\cr&&+\frac{g^{2}_{3}-g^{2}_{2}}{2g\nu}(|01\rangle|010\rangle + |10\rangle|010\rangle)
    \cr\cr&&-\frac{\sqrt{g^{2}_{1}+g_{3}^{2}}}{\sqrt{2}g}(|01\rangle|001\rangle + |10\rangle|100\rangle)\cr\cr&&+(|0e\rangle|000\rangle + |e0\rangle|000\rangle)\Big]/\frac{\sqrt{2(g_{3}^{4}-g^{2}g_{3}^{2}+2g_{3}^{2}\nu^{2})}}{g\nu},
    \cr\cr|S_{3}\rangle&\equiv&\Big[\frac{\sqrt{g^{2}_{1}+g^{2}_{3}}~(g^{2}-g^{2}_{3})}{2\sqrt{2}g\nu^{2}}
    (|01\rangle|100\rangle-|10\rangle|001\rangle)\cr\cr&&+\frac{g^{2}_{3}-g^{2}_{2}}{2g\nu}(|01\rangle|010\rangle - |10\rangle|010\rangle)
    \cr\cr&&-\frac{\sqrt{g^{2}_{1}+g_{3}^{2}}}{\sqrt{2}g}(|01\rangle|001\rangle - |10\rangle|100\rangle)\cr\cr&&+(|0e\rangle|000\rangle - |e0\rangle|000\rangle)\Big]/\frac{\sqrt{2(g_{3}^{4}-g^{2}g_{3}^{2}+2g_{3}^{2}\nu^{2})}}{g\nu},
    \cr\cr|T_{4}\rangle&\equiv&\Big[\frac{\sqrt{g^{2}_{1}+g^{2}_{3}}~(g^{2}_{3}-g^{2})}{2\sqrt{2}g\nu^{2}}
    (|01\rangle|100\rangle+|10\rangle|001\rangle)\cr\cr&&+\frac{g^{2}_{3}-g^{2}_{2}}{2g\nu}(|01\rangle|010\rangle + |10\rangle|010\rangle)
    \cr\cr&&+\frac{\sqrt{g^{2}_{1}+g_{3}^{2}}}{\sqrt{2}g}(|01\rangle|001\rangle + |10\rangle|100\rangle)\cr\cr&&+(|0e\rangle|000\rangle + |e0\rangle|000\rangle)\Big]/\frac{\sqrt{2(g_{3}^{4}-g^{2}g_{3}^{2}+2g_{3}^{2}\nu^{2})}}{g\nu},
    \cr\cr|S_{4}\rangle&\equiv&\Big[\frac{\sqrt{g^{2}_{1}+g^{2}_{3}}~(g^{2}_{3}-g^{2})}{2\sqrt{2}g\nu^{2}}
    (|01\rangle|100\rangle-|10\rangle|001\rangle)\cr\cr&&+\frac{g^{2}_{3}-g^{2}_{2}}{2g\nu}(|01\rangle|010\rangle - |10\rangle|010\rangle)
    \cr\cr&&+\frac{\sqrt{g^{2}_{1}+g_{3}^{2}}}{\sqrt{2}g}(|01\rangle|001\rangle - |10\rangle|100\rangle)\cr\cr&&+(|0e\rangle|000\rangle - |e0\rangle|000\rangle)\Big]/\frac{\sqrt{2(g_{3}^{4}-g^{2}g_{3}^{2}+2g_{3}^{2}\nu^{2})}}{g\nu},
\end{eqnarray}
where $g_{1}^{2}=g^{2}+2\nu^{2}$, $g_{2}^{2}=g^{2}-2\nu^{2}$ and $g_{3}^{2}=\sqrt{g^{4}+4\nu^{4}}$.

\begin{center}
{\bf{ACKNOWLEDGMENT}}
\end{center}
Shi-Lei Su thank Li-Tuo Shen for helpful discussions. We also
would like to thank the anonymous reviewers for their
constructive comments that helped in improving the quality of this paper.
This work was supported by the National Natural Science Foundation
of China under Grant Nos. 11264042, 11465020, 11204028 and 61465013.

\end{document}